\documentclass[11pt, conference, letterpaper, 11pt]{IEEEtran}
\IEEEoverridecommandlockouts

\usepackage{cite}
\usepackage{amsmath,amssymb,amsfonts}
\usepackage{algorithmic}
\usepackage{graphicx}
\usepackage{textcomp}
\usepackage{xcolor}
\usepackage{placeins}   

\def\BibTeX{{\rm B\kern-.05em{\sc i\kern-.025em b}\kern-.08em
    T\kern-.1667em\lower.7ex\hbox{E}\kern-.125emX}}

\begin{document}
\addtolength{\topmargin}{0.04in}
\title{Side-Channel Attack Prevention for Post Quantum Cryptography Algorithms: A Stacked Hardware Shielding Approach}


\author{
\IEEEauthorblockN{1\textsuperscript{st} Carl Benda}
\IEEEauthorblockA{
\textit{ECE, NCSU}\\
Raleigh, NC, USA \\
cmbenda@ncsu.edu}
\and
\IEEEauthorblockN{2\textsuperscript{nd} Eren Kurshan}
\IEEEauthorblockA{
\textit{Princeton Univ.}\\
Princeton, NJ, USA \\
ekurshan@princeton.edu}
\and
\IEEEauthorblockN{3\textsuperscript{rd}Aydin Aysu}
\IEEEauthorblockA{
\textit{ECE, NCSU}\\
Raleigh, NC, USA \\
aaysu@ncsu.edu}
\and
\IEEEauthorblockN{4\textsuperscript{th}Paul Franzon}
\IEEEauthorblockA{
\textit{ECE, NCSU}\\
Raleigh, NC, USA \\
Paulf@ncsu.edu}
}

\maketitle

\begin{abstract}
Despite their theoretical quantum attack resistance, practical implementations of post-quantum cryptography (PQC) algorithms remain susceptible to side-channel attacks. In recent years, researchers have demonstrated such vulnerabilities for the NIST standards such as KYBER (ML-KEM), Dilithium (ML-DSA), Falcon (FN-DSA) and SPHINCS+ (SLH-DSA). While software-based approaches have been developed,  side-channel security challenges persist. This paper proposes a hardware-based defense for side-channel resilience that uses a vertically integrated guard-layer to shield the underlying cryptographic activity. The guard layer actively introduces decoy keys and hardware activity in addition to physically shielding through stacking. The stacking approach can complement other security measures such as software masking and shuffling in protecting Number Theoretic Transform (NTT). We outline the construction of the approach and we demonstrate through simulations a generic attack targeting intermediate computation inside an NTT butterfly. 
\end{abstract}



\section{Introduction}

Post-quantum cryptography (PQC) provides new quantum-secure algorithms to protect critical cyber-infrastructure from future quantum attacks. Despite the PQC algorithms resilience against theoretical quantum adversaries, their physical implementations have been shown to have significant side-channel vulnerabilities. In recent years, all NIST selected PQC algorithms (ML-KEM (CRYSTALS KYBER), ML-DSA (CRYSTALS Dilithium) and FN-DSA  (Falcon)) have been demonstrated to have serious side channel vulnerabilities\cite{thenmozhi2025balancing}, \cite{ma2022vulnerable}, \cite{mujdei2024side}. Shared building blocks such as number theoretic transforms have been targets for side-channel attacks. Among such blocks, number theoretic transform (NTT) enables efficient polynomial multiplication for lattice cryptography and is a common operator for such systems, including both ML-DSA and ML-KEM. NTT frequently handles secret-related material and is thus targeted with various side-channel attacks \cite{primas2017single}. 

Due to this central role in polynomial multiplication, the NTT side-channel attacks have been a growing concern. In response to these attacks, solution approaches frequently require rigorous constant-time implementations and software masking schemes to prevent key leakage. Defensive techniques and related threats cover the full-stack from the circuit-level up to algorithmic levels.  In practice, a combination of these defenses may be beneficial to provide a defense-in-depth strategy. 
 
In this paper, we propose a \emph{complementary} stacked guard-layer to protect the underlying PQC computation from side-channel attacks.  3D stacking of the shielding guad-layer provides fine-grain integration of this layer to reduce both power and electromagnetic vulnerabilities.  This paper is organized as follows: Section II provides an overview of the side-channel vulnerabilities and the related work on defense techniques; Section III
presents the proposed hardware shielding approach; Section IV provides experimental analysis and results discussions; and Section V presents concluding remarks and future work. 


\section{Background Post Quantum Cryptography Security}

\subsection{Side Channel Vulnerabilities}
Modern cybersecurity systems are shifting away from quantum-vulnerable traditional cryptographic algorithms like elliptic curve cryptography (ECC) and Rivest-Shamir-Adleman (RSA) toward quantum-resistant Post-Quantum Cryptography (PQC) \cite{joseph2022transitioning} \cite{kumar2020post}. Although post-quantum algorithms provide significantly better algorithmic security against quantum attackers, their implementations may introduce physical security challenges such as  side-channel attacks 

In conventional cryptosystems such as AES and RSA, side-channel attacks often exploit leakage from secret-dependent intermediate computations. Meanwhile, attacks against lattice-based post-quantum cryptographic algorithms can target the underlying polynomial and modular arithmetic operations (polynomial multiplication, to recover secret-key information \cite{bock2024breaking}, \cite{fan2025multivariate}). Standard PQC algorithms like ML-KEM and ML-DSA rely on common building blocks like number theoretic transform (NTT).  Such common blocks have become frequent targets for electromagnetic and power attacks, threatening the security of the private keys \cite{kumar2020post}. 

\subsection{Related Work: Full-Stack Side-Channel Protections}
Side-channel defenses operate at several complementary layers, including the representation of sensitive data, the scheduling of computations, the physical implementation, and the management of cryptographic keys. 

\subsubsection{Algorithmic Level}
At the algorithmic level, masking distributes sensitive values among randomized shares and performs computations without reconstructing the underlying secrets. For lattice-based cryptography, the linear structure of polynomial arithmetic facilitates masking, while nonlinear operations such as compression, decoding, and ciphertext comparison require specialized techniques. Bos et al. demonstrate first- and higher-order masked implementations of KYBER, including protected compression and comparison operations~\cite{bos2021masking}. The protection depends on the masking order and on preserving the assumptions of the leakage model in the implementation.

\subsubsection{Execution Level}
At the execution level, shuffling randomizes the order of independent operations to obscure the correspondence between intermediate values and trace samples. Lattice implementations offer substantial opportunities for this approach because many coefficient operations and NTT butterflies can be reordered while respecting their dependencies. Xu et al. develop a hardware-friendly Fisher--Yates shuffling architecture for KYBER and evaluate the protected FPGA implementation using correlation analysis and leakage testing~\cite{xu2025shuffling}. Shuffling changes when sensitive computations occur, whereas masking changes the values they process. The two can therefore be combined. However, shuffling does not eliminate the underlying data-dependent leakage; its effectiveness also depends on preventing an adversary from recovering the execution order or otherwise compensating for the permutation.

\subsubsection{Circuit and Architecture Levels}
At the circuit and hardware-architecture levels, hiding techniques alter the observable physical activity to make sensitive computations more difficult to distinguish. Hiding is a broad category that includes temporal randomization and the introduction of additional switching activity, and thus overlaps with execution-level defenses. A concrete lattice-based example is the KYBER FPGA implementation of Moraitis et al., which combines circuit duplication with clock randomization~\cite{moraitis2024securing}. This approach modifies the hardware execution environment to obscure leakage from the cryptographic computation. Such techniques incur hardware and energy costs and require evaluation under the intended measurement conditions. Consequently, hardware hiding and algorithmic masking provide complementary forms of protection, rather than interchangeable security guarantees.

\subsubsection{Protocol and Key Management Level}
At the protocol and key-management levels, fresh re-keying limits the repeated use of an operational key, reducing opportunities to accumulate exploitable leakage about that key. The long-term secret is instead processed by a rekeying function designed to be easier to protect. Dziembowski et al. provide a lattice-based example through a construction based on Learning With Rounding (LWR), alongside a separate construction based on Learning Parity with Leakage~\cite{dziembowski2016rekeying}. Their work connects fresh rekeying with functions that support efficient masking. Rekeying therefore redistributes the protection requirements: the derivation function must withstand repeated observations, and computations using fresh keys still need protection against attacks that succeed within a single execution. The LWR construction supplies a lattice-based foundation for rekeying; it is not a demonstrated rekeying modification of ML-KEM.

\subsubsection{Leakage Resilient Cryptography}
Leakage-resilient cryptography addresses the construction and security-model level by establishing security even when an adversary obtains specified information about secret state. For example, Dachman-Soled et al. analyze Ring-LWE-based Dual-Regev encryption and prove security for three classes of non-adaptive leakage~\cite{dachman2017leakage}. Such results depend on explicit restrictions on leakage and should not be interpreted as unconditional resistance to physical measurements. A related approach is to design cryptographic schemes around the cost of protecting their implementations. POLKA, a lattice-based public-key encryption scheme, combines randomized decryption using a dummy ciphertext, efficiently maskable computations, and hard physical learning problems to reduce that cost~\cite{hoffmann2023polka}. These examples illustrate how leakage considerations can influence the cryptographic construction itself, beyond adding countermeasures to an existing implementation.

\begin{figure}[t]
  \renewcommand{\thefigure}{\arabic{figure}a}
  \centerline{\includegraphics[width=\columnwidth]{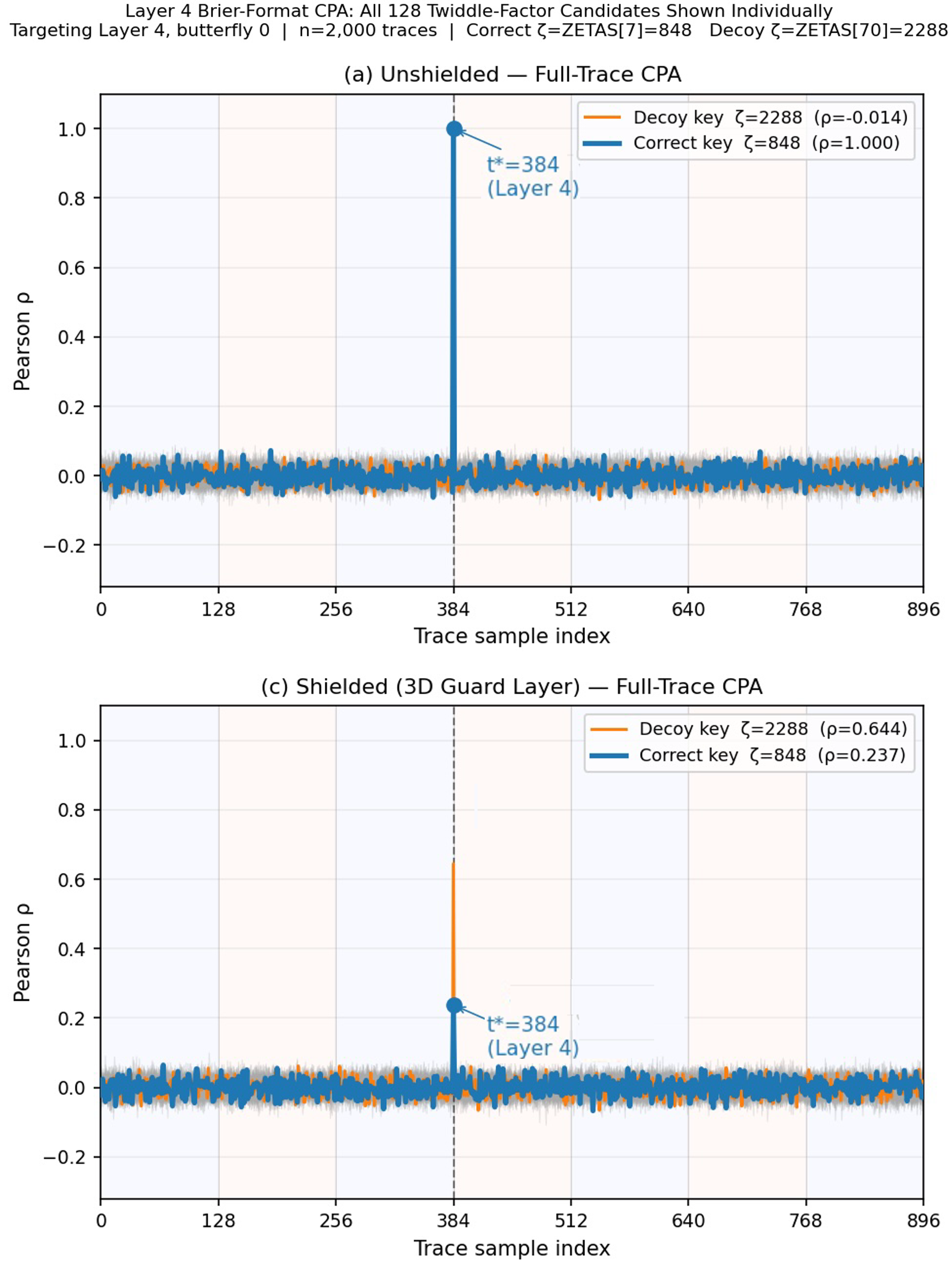}}
  \caption{Layer 4 Brier-format CPA full-trace results ($n=2{,}000$ traces,
    all 128 twiddle-factor candidates shown individually in gray).
    \textbf{(a) Unshielded:} correct key $\mathtt{ZETAS}[7]{=}848$ peaks
    sharply at $\rho=1.000$ at $t^*=384$ (Layer~4); all other candidates
    remain near zero. \textbf{(c) Shielded (3D Guard Layer):} the decoy
    $\mathtt{ZETAS}[70]{=}2288$ becomes the apparent winner ($\rho=0.644$),
    suppressing the true key to $\rho=0.237$.}
  \label{fig:cpa_traces}
\end{figure}

\begin{figure}[t]
  \addtocounter{figure}{-1}
  \renewcommand{\thefigure}{\arabic{figure}b}
  \centerline{\includegraphics[width=\columnwidth]{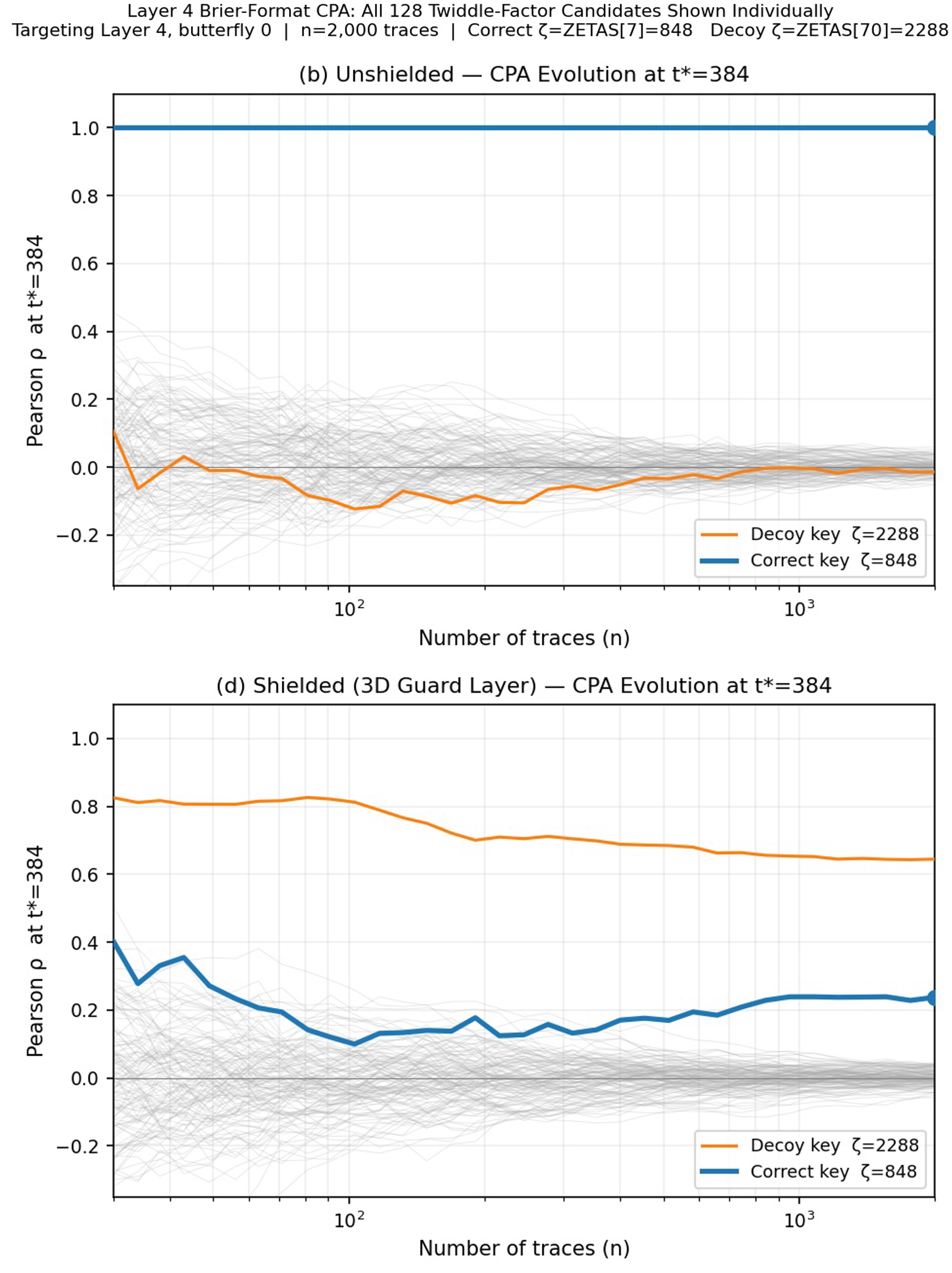}}
  \caption{CPA convergence at $t^*=384$ vs.\ number of traces (all 128
    candidates in gray). \textbf{(b) Unshielded:} correct key
    $\mathtt{ZETAS}[7]{=}848$ reaches $\rho=1.0$ from the very first traces
    and holds stably. \textbf{(d) Shielded (3D Guard Layer):} the decoy
    $\mathtt{ZETAS}[70]{=}2288$ converges to $\rho=0.644$ while the true key
    is suppressed to $\rho=0.237$.}
  \label{fig:cpa_evolution}
\end{figure}
\renewcommand{\thefigure}{\arabic{figure}}


\section {3D Guard-Layer: Hardware Shielding Approach}
This paper proposes a hardware shielding approach, which introduces an additional, \emph{compelementary} defense at the physical-integration layer. A guard layer die executes decoy computations intended to produce prominent leakage associated with incorrect hypotheses \cite{kurshan20253d}, thereby misleading analysis of the underlying lattice-cryptographic engine. 

Its distinguishing objective is to create structured, misleading observations, in addition to reducing the visibility of genuine activity. This objective differs from POLKA's use of a dummy ciphertext to randomize internal decryption computations and from duplication with clock randomization to obscure hardware activity~\cite{moraitis2024securing,hoffmann2023polka}.As protected cryptographic layer and the guard-layer share a fine-grain vertical power grid and are vertically adjacent (which helps with EM shielding as well), it becomes significantly difficult to separate the shielding computation from the actual cryptographic activity.  We assume a non-adaptive external power measurement adversary without prior knowledge of the guard layer's decoy parameters. The stacked layer could complement masking, shuffling, and rekeying while preserving the functional computation on the protected die. Its effectiveness must be assessed against adversaries that recognize the decoy mechanism and attempt to separate its contribution from the remaining secret-dependent leakage.



\begin{figure}[t]
  \centerline{\includegraphics[width=\columnwidth]{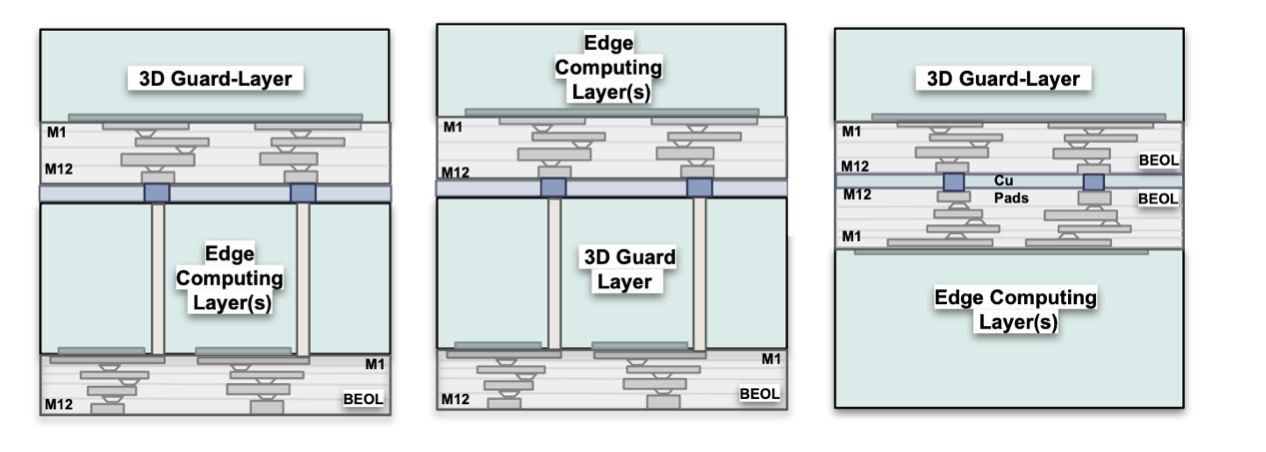}}
  \caption{ (i) Face-to-Back 3D Guard-layer configuration with PQC layer codesign (ii) Face-to-Back configuration with limited changes to PQC layer (iii) Face-to-Face Configuration}
  \label{fig:fig3}
\end{figure}

As shown in Fig.~\ref{fig:fig3} stacked security layer can be incorporated with the post-quantum computing logic through face-to-face or face-to-back 3D integration. The 3D stacking configuration provides tightly coupled power delivery network as well as physical shielding of the underlying device layers from EM antennas. 3D Integration has been shown to enhance protection against side-channel attacks in the past \cite{gu2016leveraging},\cite{gu2016thermal}, \cite{knechtel20193d}. The macro-level granularity of the 3D power grid, frequently proportional to the through-silicon-via (TSV) pitch makes it significantly difficult to separate the cryptographic activity from decoy computation.

\section{Experimental Analysis and Results}

\subsection{Experimental Setup}

We evaluated the 3D side channel shielding defense on a software simulation
of KYBER NTT execution over $\mathbb{Z}_{3329}$ ($N=256$) using a
Hamming-weight power proxy model. Correlation Power Analysis (CPA) was
applied in the Brier et al.\ (2004) \cite{brier2004correlation} full-trace format, targeting
Layer 4 butterfly 0 (trace samples $t^*=384$, twiddle factor
$\mathtt{ZETAS}[7]=848$). Fig.~\ref{fig:cpa_traces} shows full-trace CPA (left column) and Fig.~\ref{fig:cpa_evolution} shows the convergence evolution (right column) for both the
unshielded and shielded configurations, with all 128 twiddle-factor candidates
shown individually. 

\subsection{Experimental Analysis}
\subsubsection{Unshielded Configuration}
In the unshielded configuration, (Fig.~\ref{fig:cpa_traces} top), CPA identifies
$\mathtt{ZETAS}[7]=848$ as the highest-correlation butterfly intermediate
with Pearson correlation $\rho=1.0$ at $t^*=384$ from the very first traces, (ideal, noise-free simulation);
all 126 remaining candidates are indistinguishable from zero. 

\subsubsection{3D Guard-Layer: Shielded Configuration}
Using the guard layer, 
(Fig.~\ref{fig:cpa_traces} bottom), a decoy core running the same input polynomial
with a hardwired wrong twiddle factor $\mathtt{ZETAS}[70]=2288$ injects a
correlated but incorrect power signal. CPA now identifies
$\mathtt{ZETAS}[70]=2288$ as the apparent winner ($\rho=0.644$) while the
original target $\mathtt{ZETAS}[7]=848$ is suppressed to $\rho=0.237$; an
attacker committing to the highest-correlation candidate would follow the
misleading decoy. The decoy twiddle factor is selected to maximize correlation of the injected signal with a plausible but incorrect hypothesis, ensuring the attacker's highest-correlation candidate resolves to the decoy rather than the true target.

\begin{figure}[t]
  \centerline{\includegraphics[width=\columnwidth]{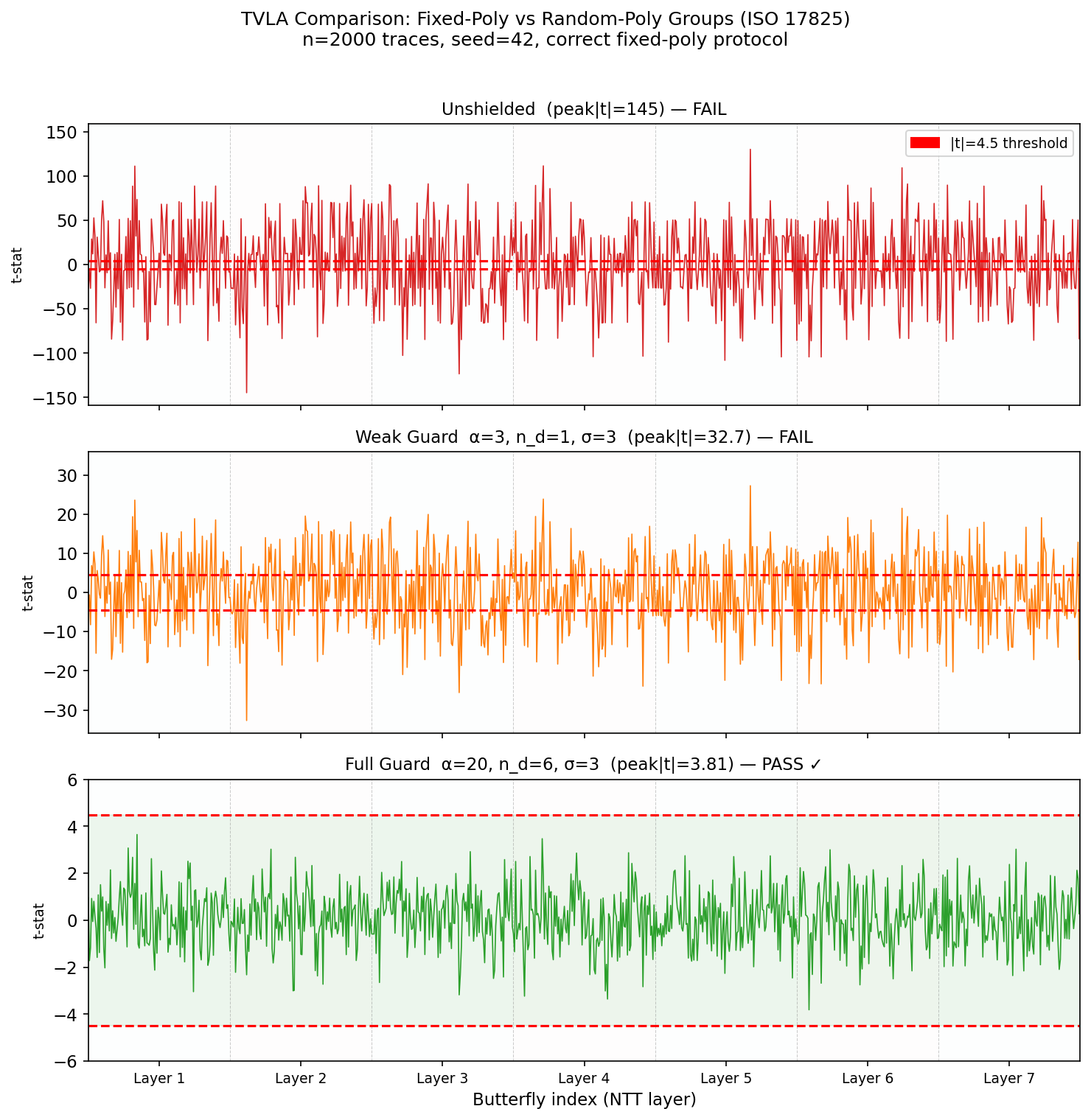}}
  \caption{TVLA comparison (ISO/IEC~17825, $n=2000$ traces). Top: unshielded
    NTT, peak $|t|=145$ --- FAIL. Middle: weak guard layer ($\alpha=3$,
    $n_d=1$), peak $|t|=32.7$ --- FAIL. Bottom: full guard layer ($\alpha=20$,
    $n_d=6$), peak $|t|=3.81 < 4.5$ --- PASS. Red dashed lines mark the
    $|t|=4.5$ detection threshold.}
  \label{fig:tvla}
\end{figure}

The evolution plot (Fig.~\ref{fig:cpa_evolution} bottom) confirms this misdirection is stable across all $n = 2{,}000$
measurements used for the TVLA test. Separately, Test Vector
Leakage Assessment (TVLA, ISO/IEC~17825) confirms that the full guard layer
(coupling $\alpha=20$, six decoy cores) passes the leakage threshold with
peak $|t|=3.81 < 4.5$ (Fig.~\ref{fig:tvla}). Fig.~\ref{fig:tvla} also
shows the progression: the unshielded NTT fails with peak $|t|=145$, a weak
guard layer ($\alpha=3$, $n_d=1$) reduces leakage to $|t|=32.7$ but still
fails, and the full configuration achieves a clean pass.

  While this study does not assume that existing multi-trace attacks extract secret data directly from butterfly operations and acknowledge that current attacks often target the coefficient-wise multiplication following the NTT where known inputs are combined with fixed secrets,  the proposed solution can be extended to help prevent single-trace attacks or the butterflies in the Inverse-NTT stage following the coefficient multiplications. The attack shown on public twiddle factors merely shows the possibility of targeting multiply-add operators inside the butterfly unit.

\section{Conclusions}


Post-quantum cryptographic implementations face  a significant side-channel vulnerability. The NTT arithmetic common to lattice-based PQC schemes creates a data-dependent power
signature exploitable by CPA.  This paper presents a 3D stacked guard layer to reduce the underlying vulnerabilities. Through the stacking and vertical integration of a shielding guard-layer die with shared power grid, the primary cryptographic layer drives a decoy core with
a hardwired wrong twiddle factor ($\mathtt{ZETAS}[70]=2288$), we suppress
the maximum-correlation CPA candidate from $\rho=1.0$ to $\rho=0.237$ and
simultaneously elevate the decoy candidate to $\rho=0.644$, inverting the
apparent attack outcome. TVLA with coupling $\alpha=20$ across six decoy cores
confirms residual leakage falls below the ISO/IEC~17825 threshold at
$|t|_{\max}=3.81 < 4.5$. The guard layer is fully transparent to the main
die with no firmware changes, no NTT modifications or functional latency overhead. This work demonstrates that 3D integration provides key security capabilities for PQC side channel attack prevention. Future work will quantify the tradeoff between decoy core count, active power overhead, and achievable CPA suppression, and evaluate physical implementation on FPGA platforms targeting the full lattice-based PQC algorithm suite.




\bibliographystyle{IEEEtran}
\bibliography{all_references.bib}

@inproceedings{brier2004correlation,
  author    = {Brier, Eric and Clavier, Christophe and Olivier, Francis},
  title     = {Correlation Power Analysis with a Leakage Model},
  booktitle = {Cryptographic Hardware and Embedded Systems -- CHES 2004},
  series    = {Lecture Notes in Computer Science},
  volume    = {3156},
  pages     = {16--29},
  publisher = {Springer},
  year      = {2004},
  doi       = {10.1007/978-3-540-28632-5_2}
}

@inproceedings{thenmozhi2025balancing,
  author    = {Thenmozhi, A. and Usha, G.},
  title     = {Balancing Quantum and Classical Security: Side-Channel Challenges
               in {NIST} {PQC} Finalist},
  booktitle = {2025 3rd International Conference on Sustainable Computing and
               Data Communication Systems (ICSCDS)},
  pages     = {2025--2031},
  publisher = {IEEE},
  year      = {2025},
  doi       = {10.1109/ICSCDS65426.2025.11167157}
}

@inproceedings{ma2022vulnerable,
  author    = {Ma, Haocheng and Pan, Shijian and Gao, Ya and He, Jiaji and
               Zhao, Yiqiang and Jin, Yier},
  title     = {Vulnerable {PQC} Against Side Channel Analysis -- A Case Study
               on {Kyber}},
  booktitle = {2022 Asian Hardware Oriented Security and Trust Symposium
               (AsianHOST)},
  pages     = {1--6},
  publisher = {IEEE},
  year      = {2022},
  doi       = {10.1109/AsianHOST56390.2022.10022165}
}

@article{mujdei2024side,
  author    = {Mujdei, Catinca and Wouters, Lennert and Karmakar, Angshuman and
               Beckers, Arthur and Bermudo Mera, Jose Maria and
               Verbauwhede, Ingrid},
  title     = {Side-Channel Analysis of Lattice-Based Post-Quantum Cryptography:
               Exploiting Polynomial Multiplication},
  journal   = {ACM Transactions on Embedded Computing Systems},
  volume    = {23},
  number    = {2},
  articleno = {27},
  numpages  = {23},
  pages     = {27:1--27:23},
  year      = {2024},
  doi       = {10.1145/3569420}
}

@inproceedings{primas2017single,
  author    = {Primas, Robert and Pessl, Peter and Mangard, Stefan},
  title     = {Single-Trace Side-Channel Attacks on Masked Lattice-Based
               Encryption},
  booktitle = {Cryptographic Hardware and Embedded Systems -- CHES 2017},
  series    = {Lecture Notes in Computer Science},
  volume    = {10529},
  pages     = {513--533},
  publisher = {Springer},
  year      = {2017},
  doi       = {10.1007/978-3-319-66787-4_25}
}

@misc{kurshan20253d,
  author        = {Kurshan, Eren and Xie, Yuan and Franzon, Paul},
  title         = {{3D} Guard-Layer: An Integrated Agentic {AI} Safety System
                   for Edge Artificial Intelligence},
  year          = {2025},
  eprint        = {2511.08842},
  archivePrefix = {arXiv},
  url           = {https://arxiv.org/abs/2511.08842}
}

@inproceedings{gu2016thermal,
  author    = {Gu, Peng and Stow, Dylan C. and Barnes, Russell and
               Kursun, Eren and Xie, Yuan},
  title     = {Thermal-Aware {3D} Design for Side-Channel Information Leakage},
  booktitle = {2016 IEEE 34th International Conference on Computer Design
               (ICCD)},
  pages     = {520--527},
  publisher = {IEEE},
  year      = {2016},
  doi       = {10.1109/ICCD.2016.7753336}
}

@inproceedings{gu2016leveraging,
  author    = {Gu, Peng and Li, Shuangchen and Stow, Dylan C. and
               Barnes, Russell and Liu, Liu and Xie, Yuan and Kursun, Eren},
  title     = {Leveraging {3D} Technologies for Hardware Security:
               Opportunities and Challenges},
  booktitle = {Proceedings of the 26th ACM Great Lakes Symposium on VLSI
               (GLSVLSI)},
  pages     = {347--352},
  publisher = {Association for Computing Machinery},
  year      = {2016},
  doi       = {10.1145/2902961.2903512}
}

@article{joseph2022transitioning,
  author  = {Joseph, David and Misoczki, Rafael and Manzano, Marc and
             Tricot, Joe and Dominguez Pinuaga, Fernando and Lacombe, Olivier and
             Leichenauer, Stefan and Hidary, Jack and Venables, Phil and
             Hansen, Royal},
  title   = {Transitioning Organizations to Post-Quantum Cryptography},
  journal = {Nature},
  volume  = {605},
  number  = {7909},
  pages   = {237--243},
  year    = {2022},
  doi     = {10.1038/s41586-022-04623-2}
}

@inproceedings{kumar2020post,
  author    = {Kumar, Manoj and Pattnaik, Pratap},
  title     = {Post Quantum Cryptography ({PQC}) -- An Overview:
               (Invited Paper)},
  booktitle = {2020 IEEE High Performance Extreme Computing Conference (HPEC)},
  pages     = {1--9},
  publisher = {IEEE},
  year      = {2020},
  doi       = {10.1109/HPEC43674.2020.9286147}
}

@inproceedings{knechtel20193d,
  author    = {Knechtel, Johann and Patnaik, Satwik and Sinanoglu, Ozgur},
  title     = {{3D} Integration: Another Dimension Toward Hardware Security},
  booktitle = {2019 IEEE 25th International Symposium on On-Line Testing and
               Robust System Design (IOLTS)},
  pages     = {147--150},
  publisher = {IEEE},
  year      = {2019},
  doi       = {10.1109/IOLTS.2019.8854395}
}

@inproceedings{bock2024breaking,
  author    = {Alpirez Bock, Estuardo and Banegas, Gustavo and
               Brzuska, Chris and Chmielewski, {\L}ukasz and
               Puniamurthy, Kirthivaasan and {\v S}orf, Milan},
  title     = {Breaking {DPA}-Protected {Kyber} via the Pair-Pointwise
               Multiplication},
  booktitle = {Applied Cryptography and Network Security -- ACNS 2024},
  series    = {Lecture Notes in Computer Science},
  volume    = {14584},
  pages     = {101--130},
  publisher = {Springer},
  year      = {2024},
  doi       = {10.1007/978-3-031-54773-7_5}
}

@article{fan2025multivariate,
  author  = {Fan, Haopeng and Zhang, Hailong and Wang, Yongjuan and
             Wang, Wenhao and Zhang, Haojin and Yuan, Qingjun},
  title   = {Multivariate Template Attack Against {NTT}-Based Polynomial
             Multiplication of {Dilithium}},
  journal = {IEEE Transactions on Information Forensics and Security},
  volume  = {20},
  pages   = {8570--8582},
  year    = {2025},
  doi     = {10.1109/TIFS.2025.3598577}
}

@article{bos2021masking,
  author  = {Bos, Joppe W. and Gourjon, Marc and Renes, Joost and
             Schneider, Tobias and van Vredendaal, Christine},
  title   = {Masking {Kyber}: First- and Higher-Order Implementations},
  journal = {IACR Transactions on Cryptographic Hardware and Embedded Systems},
  volume  = {2021},
  number  = {4},
  pages   = {173--214},
  year    = {2021},
  doi     = {10.46586/tches.v2021.i4.173-214}
}

@article{xu2025shuffling,
  author  = {Xu, Dejun and Wang, Kai and Tian, Jing},
  title   = {A Hardware-Friendly Shuffling Countermeasure Against Side-Channel
             Attacks for {Kyber}},
  journal = {IEEE Transactions on Circuits and Systems II: Express Briefs},
  volume  = {72},
  number  = {3},
  pages   = {504--508},
  month   = mar,
  year    = {2025},
  doi     = {10.1109/TCSII.2025.3528751}
}

@article{moraitis2024securing,
  author  = {Moraitis, Michail and Ji, Yanning and Brisfors, Martin and
             Dubrova, Elena and Lindskog, Niklas and Englund, H{\aa}kan},
  title   = {Securing {CRYSTALS-Kyber} in {FPGA} Using Duplication and
             Clock Randomization},
  journal = {IEEE Design \& Test},
  volume  = {41},
  number  = {5},
  pages   = {7--16},
  year    = {2024},
  doi     = {10.1109/MDAT.2023.3298805}
}

@inproceedings{dziembowski2016rekeying,
  author    = {Dziembowski, Stefan and Faust, Sebastian and Herold, Gottfried and
               Journault, Anthony and Masny, Daniel and
               Standaert, Fran{\c c}ois-Xavier},
  title     = {Towards Sound Fresh Re-Keying with Hard (Physical) Learning
               Problems},
  booktitle = {Advances in Cryptology -- CRYPTO 2016},
  series    = {Lecture Notes in Computer Science},
  volume    = {9815},
  pages     = {272--301},
  publisher = {Springer},
  year      = {2016},
  doi       = {10.1007/978-3-662-53008-5_10}
}

@misc{dachman2017leakage,
  author       = {Dachman-Soled, Dana and Gong, Huijing and Kulkarni, Mukul and
                  Shahverdi, Aria},
  title        = {On the Leakage Resilience of {Ring-LWE} Based Public Key
                  Encryption},
  howpublished = {Cryptology ePrint Archive, Paper 2017/1127},
  year         = {2017},
  url          = {https://eprint.iacr.org/2017/1127}
}

@inproceedings{hoffmann2023polka,
  author    = {Hoffmann, Cl{\'e}ment and Libert, Beno{\^\i}t and
               Momin, Charles and Peters, Thomas and
               Standaert, Fran{\c c}ois-Xavier},
  title     = {{POLKA}: Towards Leakage-Resistant Post-Quantum {CCA}-Secure
               Public Key Encryption},
  booktitle = {Public-Key Cryptography -- PKC 2023},
  series    = {Lecture Notes in Computer Science},
  volume    = {13940},
  pages     = {114--144},
  publisher = {Springer},
  year      = {2023},
  doi       = {10.1007/978-3-031-31368-4_5}
}
\vspace{12pt}
\color{red}

\end{document}